\documentstyle[aps,12pt]{revtex}

\begin{document}
\title{ELECTRONIC AND MAGNETIC PROPERTIES OF FeBr$_2$ }
\author{Z.Ropka}
\address{C{enter for Solid State Physics, \'{S}w. Filip 5, 31-150 Krak\'{o}w, POLAND. 
}}
\author{R. Michalski}
\address{{Center for Solid State Physics, \'{S}w. Filip 5, 31-150 Krak\'{o}w.}\\
{Inst. of Physics, Pedagogical University, 30-084 Krak\'{o}w, POLAND.}}
\author{R. J.Radwanski}
\address{{Center for Solid State Physics, \'{S}w. Filip 5, 31-150 Krak\'{o}w.}\\
{Inst. of Physics, Pedagogical University, 30-084 Krak\'{o}w, POLAND.}}
\maketitle

\begin{abstract}
Electronic and magnetic (e-m) properties of FeBr$_2$ have been surprisingly
well described as originating from the Fe$^{2+}$ ions and their fine
electronic structure. The fine electronic structure have been evaluated
taking into account the spin-orbit (s-o) coupling, crystal-field and
inter-site spin-dependent interactions. The required magnetic doublet ground
state with an excited singlet at D=2.8 meV results from the trigonal
distortion. This effect of the trigonal distortion and a large magnetic
moment of iron, of 4.4$\mu _B$, can be theoretically derived provided the
s-o coupling is correctly taking into account. The obtained good agreement
with experimental data indicates on extremaly strong correlations of the six
3d electrons in the Fe$^{2+}$ ion yielding their full localization and the
insulating state. These calculations show that for the meaningful analysis
of e-m properties of FeBr$_2$ the spin-orbit coupling is essentially
important and that the orbital moment (0.74 $\mu _B$) is largely unquenched
(by the off-cubic trigonal distortion in the presence of the spin-orbit
coupling).
\end{abstract}


\section{Introduction}

FeBr$_2$ became famous long time ago as one of the strongest metamagnet
[1,2]. At 4.2 K an external field of 3.15 T causes a jump of the
magnetization from the almost zero value to a very big value of 110 emu/g
[1,2]. This latter value corresponds to 4.4$\mu _B$ per formula unit (f.u.).
Attributing this magnetization to the iron ions only one comes to the same
value for one Fe atom. As this value exceeds a theoretical value of 4.0$\mu
_B$, expecting for the spin-only moment, explanation for the origin of this
large magnetic moment is a challenge for the 3d-magnetism theoretisians. FeBr%
$_2$ exhibits the antiferromagnetic ordering below T$_N$ of 14.2 K. The
appearance of the AF ordering is marked in the specific heat as very
pronounced $\lambda $-type of peak [3,4]. The appearance of the AF state is
also seen in the temperature dependence of the magnetic susceptibility
[5,6]. Properties of FeBr$_2$ have been tried to be described using
effective Hamiltonians like the S$_{eff}$=1/2 Ising Hamiltonian [7] or S$%
_{eff}$=1 model with the anisotropic exchange [7,8]. Although those
Hamiltonians had the intention to describe properties of the Fe$^{2+}$ ion
such the effective description with fictitious S=1/2 or S=1 is largely
invalid owing to the fact that the Hund's rules give for the Fe$^{2+}$ ion
the real spin quantum number S=2 [9,10]. Moreover, the neglect of the
orbital contribution, due to the so-called quenching of the orbital moment,
in these effective approaches to the 3d-magnetic ions is not any more
justified. Thus we have performed exact calculations of the Fe$^{2+}$-ion
electronic structure taking into account the spin and orbital electronic
phenomena. In fact, we have used the well-known Hamiltonian for the 3d ion

\begin{equation}
H_d=H_{CF}+H_{s-o}+H_{ex}  \eqnum{1}
\end{equation}

{\bf \ but we have performed exact quasi-atomic calculations in the }$\left| 
\text{LSL}_z\text{S}_z\right\rangle ${\bf \ basis}. In this paper we present
results of calculations of electronic and magnetic (e-m) properties of FeBr$%
_2$ with the aim of providing the consistent physical description of
zero-temperature properties (e.g. the magnetic-moment value and its
direction, the insulating state) as well as temperature dependence of the
specific heat and the paramagnetic susceptibility. It turns out that e-m
properties of FeBr$_2$ are predominantly determined by the low-energy
electronic structure of the Fe$^{2+}$ ions. The Fe$^{2+}$ ion is considered
to form the 3$d^6$ system described by S=2 and L=2. The low-energy
electronic structure results from the removal of the 25-fold degeneracy. It
turns out that there are 15 localized states within 80 meV, 10 others lying
more than 2 eV above. The existence of this fine electronic structure causes
strong thermal effects in the temperature course of the specific heat and
the magnetic susceptibility. In fact, we have performed quite similar
calculations to those that have been successfully made for description of
rare-earth systems like ErNi$_5$ [11] or NpGa$_2$ [12]. Of course, the
electronic states of the 3$d$ ion are calculated in the completely different
way owing to much smaller spin-orbit coupling in 3$d$ ions.

\section{ Theoretical outline}

\subsection{The individualized-electron model analysis}

The model applied to ErNi$_5$ or NpGa$_2$ can be called an
individualized-electron model in analogy to the band(itinerant)-electron and
localized-electron model. By this name it is pointed out that atoms in the
solid preserve much of their atomic individual properties and we have to
analyze individual electronic states. For FeBr$_2$ the starting analysis of
experimental data within the individualized-electron model can be performed
as follows:

1. from the insulating state of FeBr$_2$ we infer the electronic
configuration Fe$^{2+}$Br$_2^{1-}$;

2. the opposite ionic charges assure the global stability of the compound
due to the conventional Coulomb interactions;

3. the Br$^{1-}$ ion is electronically, below 3-4 eV, and magnetically
silent as having all full electronic shells of $^{36}$Kr; the $^{36}$Kr-like
shell shows weak diamagnetism with the negligibly small value of 5$\cdot $10$%
^{-5}\mu _B$/T,

4. thus, the low-energy electronic structure and the magnetism of FeBr$_2$
is attributed to the single-ion Fe$^{2+}$ states.

5. FeBr$_2$ has a yellow colour - it points to the existence of an energy
separation of 2.0-2.2 eV;

6. FeBr$_2$ exhibits the antiferromagnetic order below T$_N$ of 14.2 K,

7. from the crystallographic structure it is known that the Fe ion in FeBr$%
_2 $ is surrounded by 6 Br ions [4,6,8]; the Br ions form the almost cubic
octahedral surrounding (this fact can be somehow surprising owing to the
hexagonal elementary cell);

8. from detailed crystallographic considerations we note that this
octahedral surrounding in the hexagonal unit cell can be easily distorted
along the local cube diagonal; in the hexagonal unit cell this local cube
diagonal lies along the hexagonal c axis and the related distortion can be
described as the trigonal distortion.

\subsection{The Fe$^{2+}$ ion}

The Fe$^{2+}$ ion has 6 3$d$ electrons. In our approach they form the
highly-correlated $d^6$ electronic system fulfilling the Russel-Saunders
LS-coupling scheme with the ground term described by Hund's rules. Such the
approach is different from the present band structure theories that
consider, at least at the start, these 6 electrons as largely independent
(the one-electron approach) introducing later (strong) electron
correlations. Thus our treatment starts from the opposite limit, i.e. from
the highly-correlated electrons within the incomplete 3$d$ shell. The
comparison with experiments should verify these approaches. Results of our
calculations indicate the significant physical adequancy of our approach.

\subsection{The electronic structure of the Fe$^{2+}$ ion as the highly
correlated 3$d^6$ system.}

The 3$d^6$ system, according to Hund's rules, is described by S=2 and L=2
and the ground term is $^5$D. This term is 25-fold degenerated (see ref.
[13], for instance). The 25-fold degeneracy is removed by:

1. the cubic crystal-field, ($H_{CF}^{cub}$)

2. the spin-orbit coupling ($H_{s-o}=\lambda L\cdot S$)

3. the trigonal-lattice distortion ( B$_2^0O_2^0$ ).

These interactions have been written in the decreasing energy sequence.

The 25 levels and their eigenfunctions have been calculated by the direct
diagonalization of the Hamiltonian (1) within the $|LSL_zS_z>$ base. It
takes a form [14]:

\begin{equation}
H_d=H_{cub}+\lambda L\cdot S+B_0^2O_0^2+\mu _B(L+g_sS)\cdot B_{mol} 
\eqnum{2}
\end{equation}

The separation of the crystal-electric-field (CEF) Hamiltonian for the cubic
and off-cubic part is made for the illustration reason as the cubic crystal
field is usually very predominant. The obtained energy level scheme is shown
in Fig. 1.

The cubic CEF Hamiltonian takes, for the z axis along the cube diagonal, the
form

\begin{equation}
H_{cub}=-\text{ }\frac 23B_4\cdot (O_4^0+20\sqrt{2}O_4^3)  \eqnum{3}
\end{equation}

$O_m^n$ are the Stevens operators. These three above-mentioned interactions
yield a magnetic doublet ground state with an excited singlet at 33 K, see
Fig.1e. The degeneracy of the ground doublet is spontaneously removed by the
formation of the magnetic order. Then in calculations a Zeeman-like
Hamiltonian 
\begin{equation}
H_z=-\mu _{\text{B}}(L+g_sS)\cdot B_{mol}  \eqnum{4}
\end{equation}

with $g_s$=2.0023 and the molecular field B$_{mol}$ appears.

\subsection{The inter-site spin interactions in FeBr$_2$}

In the ordered state occurring below 14.2 K the molecular field is set up
self-consistently as $B_{mol}=-n<m_d>$, where n is the molecular-field
coefficient and $m_d=-\mu _B(L+g_sS)$. It originates from inter-site spin
interactions. Thus our full Hamiltonian for FeBr$_2$ contains two, intra-
and inter-ion, terms:

\begin{equation}
H=\sum H_d+\sum H_{d-d}  \eqnum{5}
\end{equation}

where summation goes over all d ions. $H_{d-d}$ originates from the
Zeeman-like term and is written as

\begin{equation}
H_{d-d}=\sum \sum n\cdot (m_d\cdot <m_d>-\frac 12<m_d>^2)  \eqnum{6}
\end{equation}

The last term in (6) is included in order to avoid the double counting (see,
for instance refs 11,12). The summation goes over all pairs of magnetic
ions. This inter-site interactions produce the magnetic state; in case of
FeBr$_2$ it is antiferromagnetic arrangement (along the c axis) of the
ferromagnetic planes ($\perp c$).

\subsection{The specific heat of FeBr$_2$}

The d-electron specific heat is calculated [15] by making use of the general
formula [11,12]:

\begin{equation}
c_d(T)=-T\frac{\delta ^2F(T)}{\delta T^2}  \eqnum{7}
\end{equation}

where $F(T)$ is the free energy calculated over the available energy states
resulting from the consideration of the Hamiltonian (2), with (6) if the
magnetically-ordered state is realized.

The paramagnetic susceptibility $\chi _d(T)$ is evaluated as the induced
magnetization by the applied field, say of 1T, provided that the
magnetization is linear with the field as is in the present case.

\section{Results and discussion}

Results for the temperature dependence of the specific heat c$_d$(T) and of
the magnetic susceptibility $\chi _d(T)$ of the Fe$^{2+}$-ion system are
shown in Figs 2 and 3, respectively. The parameters used are:

1. the cubic $B_4$ parameter is taken as +200 K in order to get the overall
crystal-field splitting E$_g$-T$_{2g}$ of about 2 eV (=120$\cdot B_4$) -
this is for the explanation of the yellow colour of FeBr$_2$,

2. the spin-orbit coupling $\lambda $ equals -150 K as is given for the Fe$%
^{2+}$ ion in ref. 13 on p. 399.

3. the trigonal distortion B$_2^0$=-30 K yields the singlet-doublet
splitting of D=33 K (2.8 meV) with the doublet lower, see Fig. 1e,

4. the molecular-field coefficient n= -0.8 K/$\mu _B^2$ (= -1.2 T/$\mu _B$)
has been adjusted in order to reproduce the experimentally observed Neel
temperature of 14.2 K.

All these parameters are physically very reasonable. They provide the
electronic structure of the Fe$^{2+}$ ion as is shown in Fig. 1. The cubic
CEF is responsible for the large energy scale (say, 1.5-4.0 eV) and the s-o
coupling for the medium, 25-200 meV, energy scale. The off-cubic distortion
causes energetical effects up to 10-15 meV. The magnetic interactions range
up to, say, 20 meV for compounds with the high magnetic ordering
temperature. In case of FeBr$_2$ they are of 1.5 meV only. The inclusion of
the relatively weak intersite magnetic $H_{d-d}$ interactions is
indispensable for the appearance of the long-range magnetic order. They
cause an internal magnetic field, of 5.15 T at 0 K, and a Zeeman-like
splitting of the doublet levels. The splitting is strongly temperature
dependent and it diminishes at T$_N$. The temperature dependence of the
three lowest levels is shown in Fig. 4. These three levels with their
temperature dependence are responsible for the low-temperature specific
heat. In fact, the $\lambda $-peak dependence is mainly determined by the
temperature dependence of the lowest level. Of course, the temperature
dependence occurs only in the magnetically-ordered state as the temperature
dependence of the CEF levels, if exist, is expected to be negligible
provided the crystallographic structure is not changed.

The magnetic-moment value is perfectly reproduced by our calculations. We
have obtained a value of 4.26 $\mu _B$ in very good agreement with
experimental data [1,2]. It is directed along the hexagonal c axis that is
the natural distortion direction (such the distortion does not change the
symmetry of the elementary cell, though it changes the local symmetry). Such
the direction assures the Ising-like behaviour of the Fe$^{2+}$ ion moment.
The present approach allows for the evaluation of the orbital and spin
contributions, Fig.5. The S$_z$ value amounts to +1.74 (the spin moment of
3.48 $\mu _B$) whereas the calculated orbital moment amounts to 0.78 $\mu _B$%
. It is very large - it amounts to 18\% of the total moment. (On the other
hand sceptists can say that it is only 39\% of the full orbital moment of 2 $%
\mu _B$.) These studies about the magnetic-moment value and the different
effects on the fine electronic structure indicate that the orbital moment
has to be ''unquenched'' in the solid-state physics. It turns out that the
trigonal distortion is much more effective in the unquenching of the orbital
momentum then the tetragonal distortion, for instance.

This large orbital contribution is also seen in the temperature dependence
of the paramagnetic susceptibility. The effective moment of 5.39 $\mu _B$,
calculated from the $\chi ^{-1}$ vs T plot at room temperature, exceeds by
11\% the spin-only moment. However, one should notice in Fig. 3 that the
general course of the calculated susceptibility differs substantially from
the S=2 plot. Our calculated moment is in very good agreement with typical
values often found in Fe$^{2+}$-ion compounds [Ref.9, Table 14.2; Ref.10,
Table 31.4]. Such the increase is attributed in the effective approaches to
the change of the g factor; instead of 2 it would be 2.20 at the present
case. Inspecting further the $\chi $(T) curve in Fig. 3 one notices that $%
\chi $(T) is not the straight line. The low-temperature behaviour indicates
on even larger value for the effective moment. Indeed, the slope in the
40-100 K range yields 6.58 $\mu _B$ pointing to the almost full orbital
contribution (6.70 $\mu _B$).

The found molecular-field coefficient n of 1.2 T/$\mu _B$, determining the
magnetic-ordering temperature, indicates that 2/3 of its is related with
antiferromagnetic interations as a value of 0.8 T/$\mu _B$ is derived from
the metamagnetic field of 3.15 T.

\section{Conclusions}

Temperature dependence of the specific heat and of the magnetic
susceptibility of FeBr$_2$ as well as the zero-temperature magnetic moment
and its direction have been surprisingly well described as originating from
low-energy localized states of the Fe$^{2+}$ ions. The present approach
provides the consistent physical description of the electronic structure in
the 0-4 eV energy range. The obtained good agreement indicates extremaly
strong correlations of the six 3d electrons in the Fe$^{2+}$ ion confirming
the validity of the Russel-Saunders scheme and Hund's rules. These
calculations show that for the meaningful analysis of electronic and
magnetic properties of FeBr$_2$ the spin-orbit coupling is essentially
important and that the orbital moment is largely ''unquenched'' (by the
off-cubic trigonal distortion in the presence of the spin-orbit coupling).
The 18\% orbital contribution proves the unapplicability of all approximate
treatments of the Hamiltonian (1) both in terms of an effective S ($\neq $2)
and a fictitious orbital L ($\neq $2) number. The calculated effective
moment of 5.39$\mu _B$ at room temperature, exceeding by 10\% the spin-only
moment, is in perfect agreement with typical values often found in Fe$^{2+}$%
-ion compounds. This remarkably good and consistent description proves a
posteriori that the fine quasi-atomic electronic structure, resulting from
the crystal-field and spin-orbit coupling, survives in the solid-state
matter.

{\bf Figure captions:}

Fig.1. The fine electronic structure of the highly-correlated 3$d^6$
electronic system. a) the 25-fold degenerated $^5$D term given by Hund's
rules: S=2 and L=2. b) the effect of the cubic octahedral crystal-field, c)
the combined action of the spin-orbit coupling and the cubic crystal field: B%
$_4$=+200K, $\lambda $= -150 K; d and e) an extra splitting produced by the
trigonal distortion - the case e is realized in FeBr$_2$; the
trigonal-distortion parameter B$_2^0$= -30K produces a spin-like gap of 2.8
meV.

Fig. 2. The calculated temperature dependence of the specific heat of FeBr$%
_2 $ (the solid line) as composed from the lattice contribution, shown by
the dashed-point line, and the d-electron contribution with the $\lambda $
peak at T$_N$. Points represent experimental data after Ref. 3.

Fig. 3. The calculated temperature dependence of the paramagnetic
susceptibility of the Fe$^{2+}$ ion in the slighly-distorted octahedral
crystal field (the solid line). The dashed line shows the susceptibility for
the exactly octahedral crystal field and the spin-orbit coupling. Points
represent experimental data for FeBr$_2$ [6, p.58].

Fig. 4. Temperature dependence of the three lowest levels showing the
splitting of the lowest doublet in the magnetic state (below T$_N$=14.2K).
The collapse of the splitting of the ground-state doublet should be noticed.

Fig.5. The calculated temperature dependence of the local magnetic moment of
the Fe$^{2+}$ ion in FeBr$_2$ together with spin and orbital contributions
at 0 K.


\begin{references}
\bibitem{15}  the computer program is available on the written request to
the author.
\end{references}
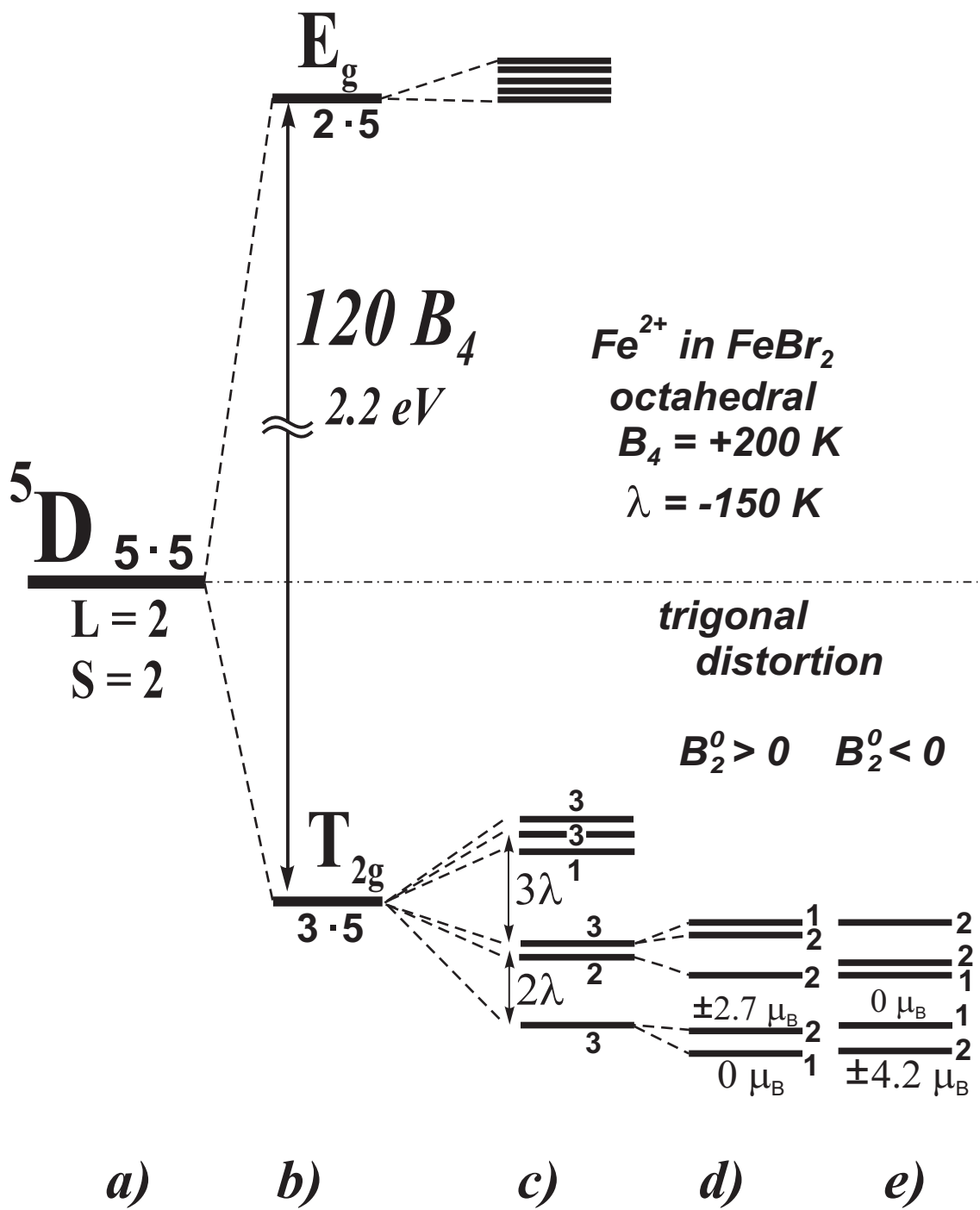
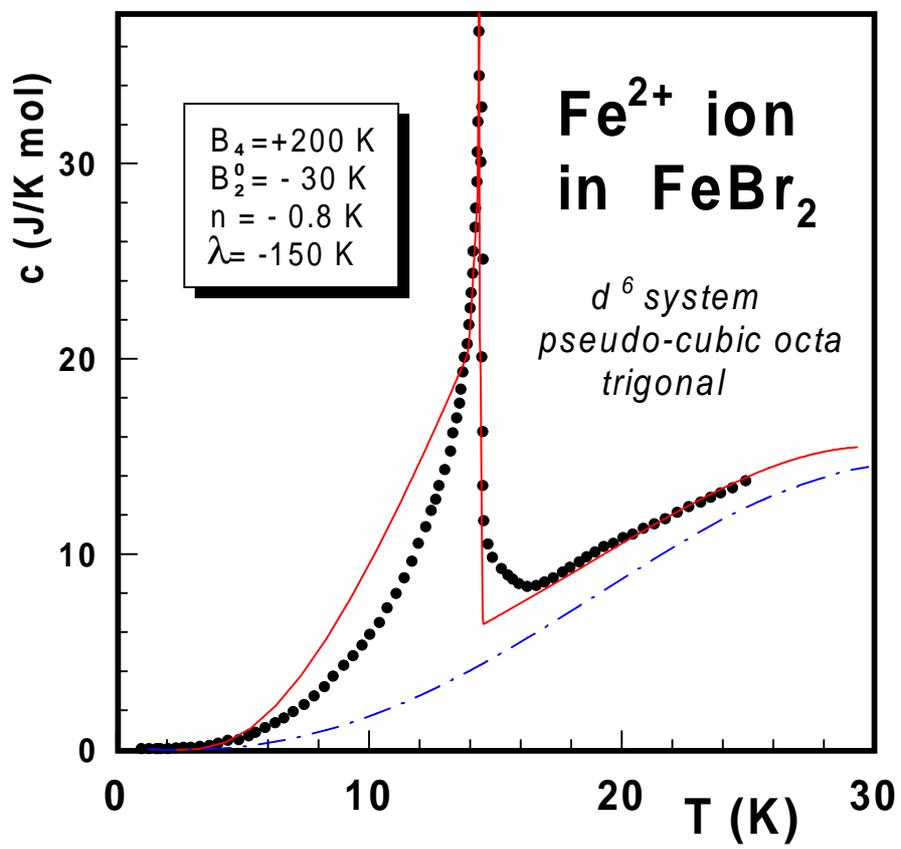
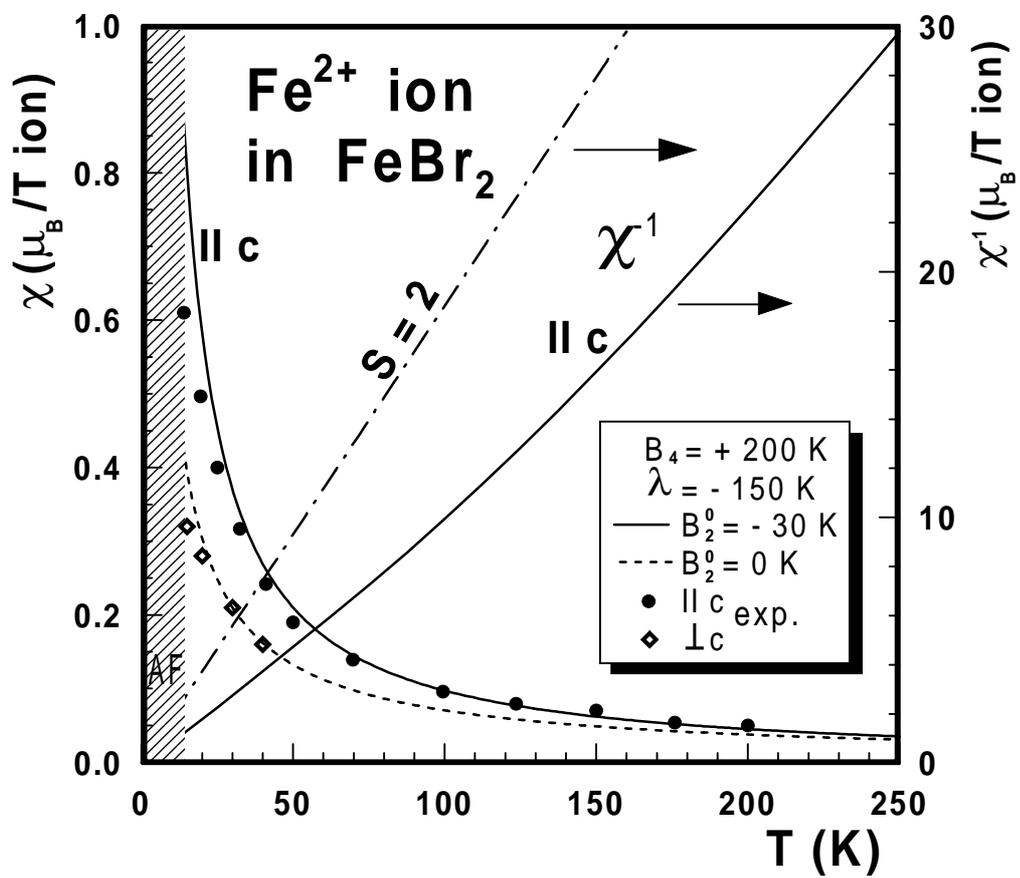
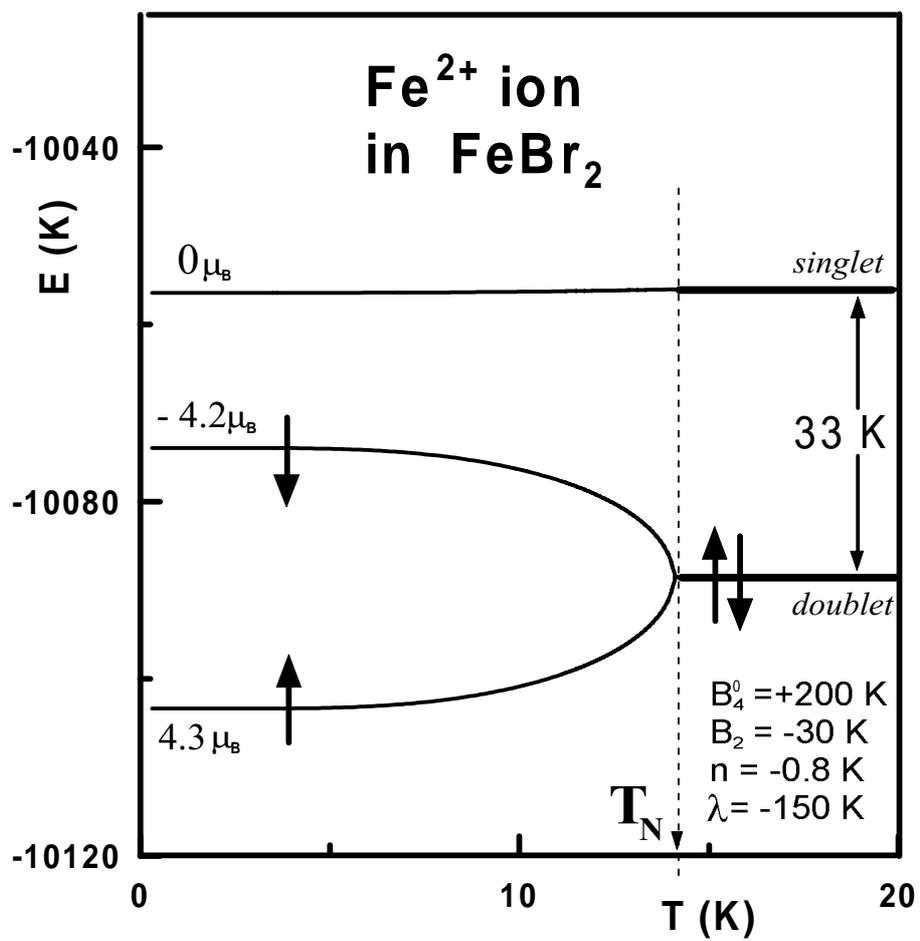
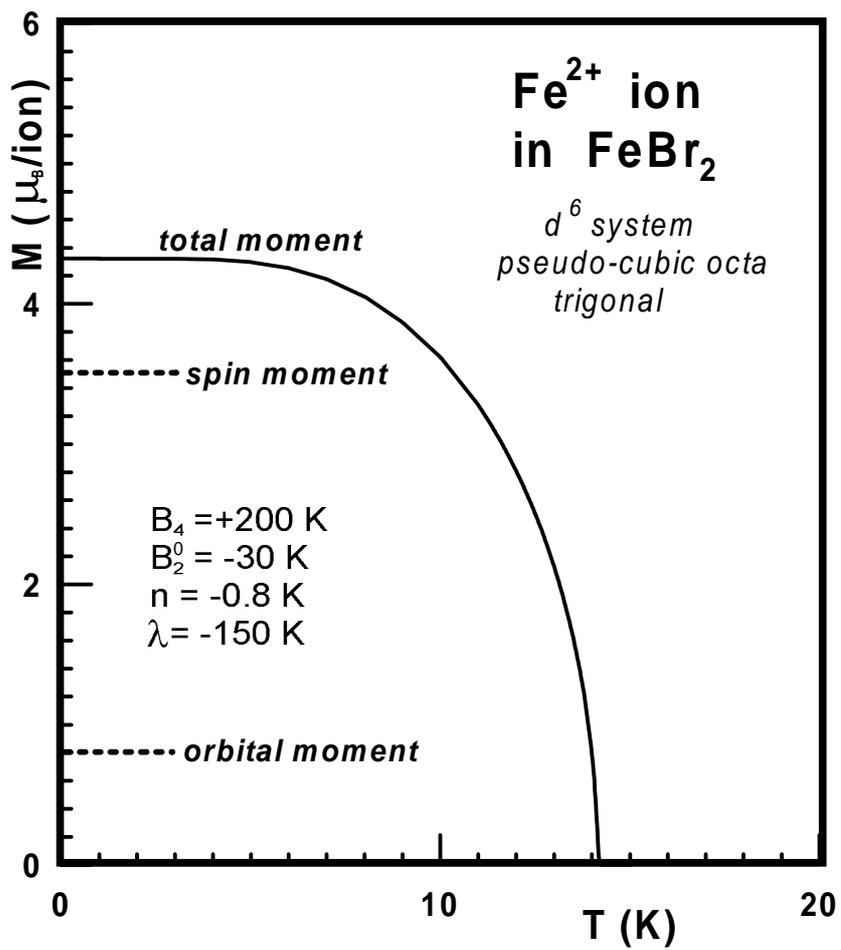
\end{document}